\documentclass[pre,showpacs,twocolumn,address]{revtex4-1}
\usepackage[dvips]{graphicx}% Include figure files
\usepackage{amsmath,times,bm}
\usepackage{soul,color}
\usepackage{appendix}
\usepackage[colorlinks,citecolor=blue,linkcolor=red]{hyperref}
\usepackage{ulem}
\usepackage{amsfonts}
\usepackage{subeqnarray}
\usepackage{cases}
\begin{document}

\title{Energy Super-Diffusion in One-Dimensional Momentum Non-Conserving Nonlinear Lattices}

\author{Hengzhe Yan}
\affiliation{Institute of Systems Science and Department of Physics, College of Information Science and Engineering, Huaqiao University, Xiamen 361021, China}
\affiliation{Center for Phononics and Thermal Energy Science, School of Physics Science and Engineering, Tongji University, 200092 Shanghai, China}

\author{Jie Ren}
\email{xonics@tongji.edu.cn}
\affiliation{Center for Phononics and Thermal Energy Science, China-EU Joint Center for Nanophononics, Shanghai Key Laboratory of Special Artificial Microstructure Materials and Technology, School of Physics Sciences and Engineering, Tongji University, Shanghai 200092, China}

\author{Nianbei Li}
\email{nbli@hqu.edu.cn}
\affiliation{Institute of Systems Science and Department of Physics, College of Information Science and Engineering, Huaqiao University, Xiamen 361021, China}

\begin{abstract}
There is a well-known mapping between energy normal (super-) diffusion and normal (anomalous) heat conduction in one-dimensional (1D) nonlinear lattices. The momentum conserving nonlinear lattices exhibit energy super-diffusion behavior with the only exception of coupled rotator model. Yet, for all other 1D momentum non-conserving nonlinear lattices studied so far, the energy diffusion or heat conduction is normal. Here we propose a 1D nonlinear lattice model with negative couplings, which is momentum non-conserving due to the translational symmetry breaking.
Our numerical results show that energy super-diffusion instead of normal diffusion can be found for this model, which indicates that neither momentum non-conservation is a sufficient condition for energy normal diffusion nor momentum conservation is a necessary condition for energy super-diffusion.
Zero frequency phonon mode at Brillouin zone boundary induces a new conserved momentum parity, which is the key for the energy super-diffusion and anomalous heat conduction. Removing the zero frequency mode, such as by on-site potential, is a sufficient condition for normal heat conduction in 1D nonlinear lattices.
\end{abstract}

\maketitle

Since the first ever discovery of anomalous heat conduction for 1D nonlinear Fermi-Pasta-Ulam $\beta$ (FPU-$\beta$) lattice \cite{Lepri1997prl}, the discussions and debates of the sufficient and necessary conditions for normal or anomalous heat conduction have never been ended \cite{Bonetto2000,Lepri2003pr,Dhar2008ap,Liu2013epjb}. In early pioneer works, anomalous heat conduction was found for momentum conserving FPU-$\beta$ \cite{Lepri1997prl} and diatomic Toda lattice \cite{Hatano1999pre} while normal heat conduction was observed for momentum nonconserving Frenkel-Kontorova (Fk) \cite{Hu1998pre} and $\phi^4$ lattices \cite{Hu2000pre,Aoki2000pla}. This stimulated the claim that momentum conservation might be the sufficient and necessary condition for anomalous heat conduction in 1D nonlinear lattices \cite{Prosen2000prl,Narayan2002prl}. However, the normal heat conduction was obtained for 1D coupled rotator model, which is a momentum conserving lattice \cite{Giardina2000prl,Gendelman2000prl}.
Recent numerical results seem suggesting that asymmetry in momentum conserving lattices can induce normal heat conduction \cite{Zhong2012pre,Savin2014pre,Chen2016jsm}, but later works demonstrate that this might be a finite size effect and anomalous heat conduction still will be approached for asymmetric momentum conserving lattices in the thermodynamical limit \cite{Wang2013pre,Das2014jsp}.

Till so far, what we can be sure of is that all the momentum nonconserving 1D nonlinear lattices with on-site potentials have been found to exhibit normal heat conduction \cite{Hu1998pre,Hu2000pre,Aoki2000pla}. It is well known that the existence of on-site potential will lift the zero phonon mode in the lattice phonon spectrum. Although introducing on-site potential will break the momentum conservation, the momentum nonconservation is not equivalent to the existence of on-site potential. Therefore, for the lattice properties of on-site potential or momentum nonconservation, it is interesting and necessary to investigate that which one of them can guarantee the normal heat conduction for 1D nonlinear lattices.

As the lattice system has no particle transport, heat conduction can be directly related to energy diffusion. It has been proved that the behavior of heat conduction has a one-to-one correspondence with the property of energy diffusion in 1D nonlinear lattice systems \cite{Liu2014prl}. The size-dependence of thermal conductivity $\kappa$ can be generally expressed as a power-law function of system length $L$ as $\kappa \propto L^{\alpha}$ \cite{Bonetto2000,Lepri2003pr,Dhar2008ap,Liu2013epjb}. The exponent $\alpha=0$ represents the normal heat conduction and $\alpha=1$ describes the ballistic heat conduction. For $0<\alpha<1$, the system exhibits the anomalous heat conduction. On the other hand, the energy diffusion can be characterized by the Mean Square Displacement (MSD) $\left<\Delta x^2(t)\right>_{E}$ of energy fluctuation. The time-dependence of energy diffusion $\left<\Delta x^2(t)\right>_{E}$ can be generally expressed as $\left<\Delta x^2(t)\right>_{E}\propto t^{\beta}$ \cite{Zhao2006prl}. The normal and ballistic energy diffusions correspond to $\beta=1$ and $\beta=2$, respectively. For $1<\beta<2$, the system exhibits anomalous super-diffusion.

The connection theory claims that $\alpha=\beta-1$ directly relating heat conduction with energy diffusion \cite{Liu2014prl}.
According to the connection theory, normal (anomalous) heat conduction corresponds to normal (anomalous) energy diffusion. This theoretical relation has been verified by numerical simulations in 1D symmetric nonlinear lattices including the FPU-$\beta$ lattice with anomalous heat conduction \cite{Zhao2006prl}, and the FK, $\phi^4$ and coupled rotator model with normal heat conduction \cite{Zhao2006prl,Li2015njp}. In particular, this relation enables us to numerically study the heat conduction problem via the energy diffusion method, which can be performed more efficiently and accurately by considering micro-canonical simulation without heat baths included.

In this paper, we propose a nonlinear lattice model without momentum conservation but sill maintaining zero frequency phonon mode. The zero frequency phonon mode is remained because this new proposed inverse-coupling model has no on-site potential. In the same time, the zero frequency phonon mode is located at the Brillouin zone boundary, not at the long-wave length limit with phonon wave-vector $k=0$ due to the breaking of momentum conservation. Therefore, this momentum nonconserving inverse-coupling model without on-site potential does possess zero frequency phonon mode, which turns out to be essential for its anomalous energy diffusion.

In the following part, the renormalized phonon dispersion relation will be theoretically developed for this inverse-coupling model. The theoretical prediction of the renormalized phonon properties will be verified by numerical simulations. We then perform detailed numerical simulations to investigate the energy diffusion behavior for this inverse-coupling model and energy super-diffusion can be observed for this momentum nonconserving model yet with zero frequency phonon mode. Our results indicate that momentum nonconservation can not guarantee normal energy diffusion or heat conduction for 1D nonlinear lattices.

\begin{figure}[t]
\includegraphics[width=1\columnwidth]{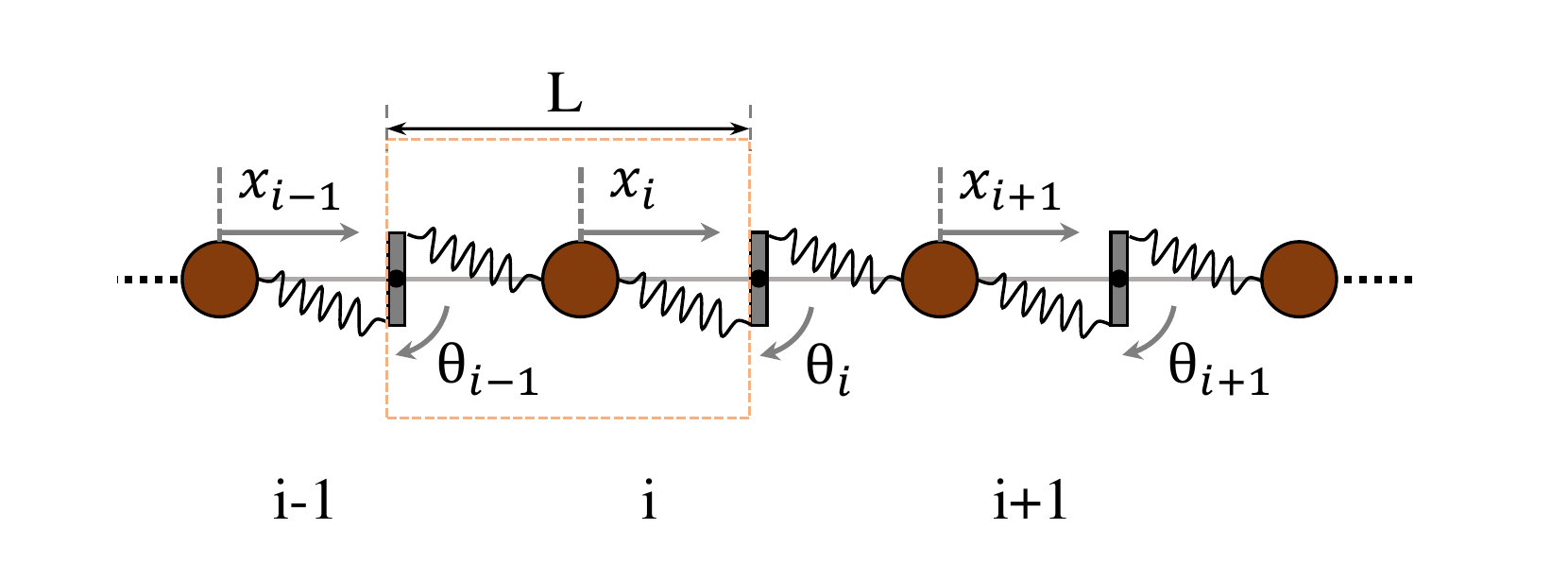}
\vspace{-0.5cm}
\caption{\label{fig:pic1}
The schematic picture of the proposed 1D inverse-coupling model. $x_i$ is the displacement from its equilibrium position for $i$-th mass. $\theta_i$ is the rotational angle for the $i$-th pole.}
\end{figure}

The inverse-coupling model is inspired by a spring-mass-pole chain as illustrated in Fig.~\ref{fig:pic1}. Each mass labeled by 'i' can move a distance $x_i$ along a certain line from its equilibrium position. Each pole can rotate an angle $\theta_i$ around its fixed center and the equilibrium orientation is vertical to the moving direction of particle. With the assumption that rotational inertia of each pole is small, the system can be reduced into one with Hamiltonian independent of $\theta_i$. Moreover, we suppose that the length of pole and moving of pole's endpoint is small compared with distance between masses. And the potential function of sprint is taken to be $V_0(y)=\gamma y^2/2+\beta y^4/4$ in analogous to FPU-$\beta$ model. The Hamiltonian then is then:
\begin{equation}\label{ic}
H=\sum_{i}\left[\frac{p_i^2}{2}+V(x_i+x_{i-1})\right]=\sum_{i}H_i,
\end{equation}
where the $V(y)=\gamma y^2/2+\beta y^4/4$ if we set $\gamma$ and $\beta$ to be some particular values. And periodic boundary condition $x_{0}=x_{N}$ is applied if total $N$ sites are considered. The detailed derivation of the Hamiltonian is shown in the appendix.

The inverse-coupling model is very similar to the FPU-$\beta$ lattice whose Hamiltonian is:
\begin{equation}
H=\sum_{i}\left[\frac{p_i^2}{2}+V(x_i-x_{i-1})\right].
\end{equation}
But for inverse-coupling model signs within interaction potential terms are positive. This difference comes from the fact that for inverse-coupling model, the increase of $x_i$ will tend to reduce the value of $x_{i-1}$ of its neighborhood, which can be seen in Fig.\ref{fig:pic1}. While for FPU-$\beta$ lattice, the increase of the displacement $x_i$ tends to increase the value $x_{i-1}$ of its neighborhood.

In order to understand the property of inverse-coupling model, we first analyze the linear inverse-coupling model with Hamiltonian:
\begin{equation}\label{ic-lin}
H=\sum_{i}\left[\frac{p_i^2}{2}+\frac{1}{2}(x_{i}+x_{i-1})^2\right]
\end{equation}
It is straightforward to derive that the total momentum $d\sum_{i}p_i/dt=-\sum_{i}(x_{i-1}+2x_i+x_{i+1})\neq 0$ is not conserved due to the lack of translational symmetry.

\begin{figure}[t]
\includegraphics[width=1\columnwidth]{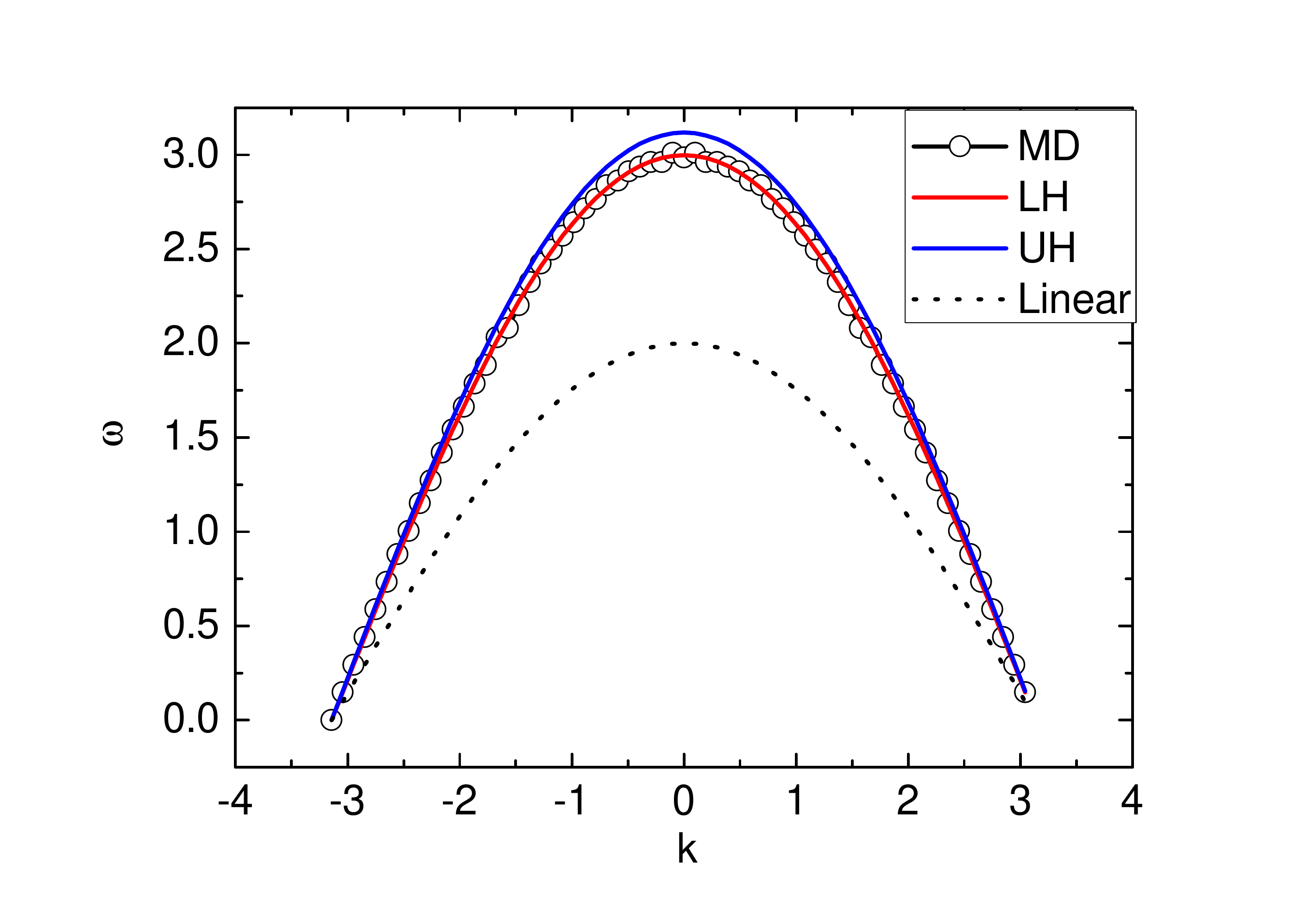}
\vspace{-0.5cm}
\caption{\label{fig:dis-fpu}
The phonon dispersion relation for inverse-coupling model. The dashed line is the analytic result for linear inverse-coupling model as $\omega_k=2\cos{(k/2)}$. The red and blue solid lines are the renormalized phonon predictions for lower limit of Eq. (\ref{eq-L-limit}) and upper limit of Eq. (\ref{eq-U-limit}), respectively. The circles are the numerical results of $\omega^R_k$ from Molecular Dynamics (MD) simulations with parameter of energy density $e=1$ corresponding to a temperature $T\approx 1.16$. The numerical $\omega^R_k$ lies between the predictions of lower limit and upper limit and close to lower limit prediction with this energy density.}
\end{figure}

The equation of motion of the linear inverse-coupling model can be obtained as $d^2x_i/dt^2=-(x_{i-1}+2x_i+x_{i+1})$, which can be solved by considering the travelling wave solution as $x_i(t)\propto e^{-j(\omega t-ki)}$ with $j$ the imaginary unit, $k$ the wave vector and $\omega$ the frequency. The dispersion relation can be derived as $\omega_k=2\cos{(k/2)},-\pi<k\leq\pi$, which is plotted as a dashed line in Fig. \ref{fig:dis-fpu}. It can be seen that $\omega_{k=0}=2$ at long-wave length limit is not a zero frequency phonon mode. However, the linear inverse-coupling model does have zero frequency phonon mode with $\omega_{k=\pi}=0$, which is shifted to the Brillouin zone boundary. This $\pi$ shift can be understood as the phase factor $e^{j\pi}=-1$ contributed by the inverse-couplings. Therefore, the breaking of translational symmetry makes the momentum not conserved any more, while the zero frequency phonon mode is maintained as a result of lacking on-site potential.

For the inverse-coupling model of Eq. (\ref{ic}) with FPU-$\beta$ like nonlinear term, a renormalized phonon dispersion relation $\omega^{R}_k$ can be derived with the renormalization phonon theory as that done for FPU-$\beta$ model \cite{Alabiso1995jsp,Alabiso2001jpa,Lepri1998pre,Nianbei2006epl,Dahai2008pre,Nianbei2012aa,Liu2015pre}. The resulted dispersion relation $\omega^{R}_k$ can be expressed as:
\begin{equation}\label{eq-omega-fpu}
\omega^R_k=\sqrt{\alpha}\omega_k=2\sqrt{\alpha}\cos{\frac{k}{2}},
\end{equation}
where the renormalization coefficient $\alpha$ is mode-independent function of the temperature $T$ due to the nonlinear interaction. According to the variational renormalization phonon theory \cite{Liu2015pre}, the coefficient $\alpha$ has a lower and upper limit expressions as $\alpha_L$ and $\alpha_U$ respectively. In particular, the coefficient $\alpha$ turns out to be the same as that for FPU-$\beta$ model as \cite{Alabiso1995jsp,Alabiso2001jpa,Lepri1998pre,Nianbei2006epl,Dahai2008pre,Nianbei2012aa,Liu2015pre}:
\begin{eqnarray}
\alpha_L&&=1+\frac{\int^{\infty}_{0}x^4 e^{-(x^2/2+x^4/4)/T}}{\int^{\infty}_{0}x^2 e^{-(x^2/2+x^4/4)/T}},\label{eq-L-limit}\\
\alpha_U&&=\frac{1}{2}\left(1+\sqrt{1+12T}\right) \label{eq-U-limit}
\end{eqnarray}
The coefficient $\alpha$ is only temperature dependent or equivalently nonlinearity dependent. The difference between two predictions of lower limit $\alpha_L$ and upper limit $\alpha_U$ are very small.

To verify the dispersion relation of Eq. (\ref{eq-omega-fpu}) in the inverse-coupling model, we apply the resonance phonon approach method to numerically calculate the renormalized phonons $\omega^R_k$ \cite{Wanglei2016pre,Wanglei2017pre}. In Fig. \ref{fig:dis-fpu}, the numerical results of renormalized phonon frequencies $\omega^R_k$ are plotted for the inverse-coupling model with energy density $e=1$ corresponding to temperature $T=1.16$. The theoretical lower limit $\alpha_L$ and upper limit $\alpha_U$ are also plotted as red and blue lines respectively for comparisons. It can be seen that the numerical results at this temperature are between the two predictions of $\alpha_L$ and $\alpha_U$ and close to the lower limit $\alpha_L$. Therefore, the dispersion relations in linear and nonlinear inverse-coupling models share the same property that the long-wave length limit phonon mode at $k=0$ does not have zero frequency. This is the result of the breaking of translational symmetry and momentum conservation. On the other hand, the zero frequency phonon mode still exists at the Brillouin zone boundary at $k=\pm\pi$ since there is no on-site potential to lift the zero frequency mode.

\begin{figure}[t]
\includegraphics[width=1\columnwidth]{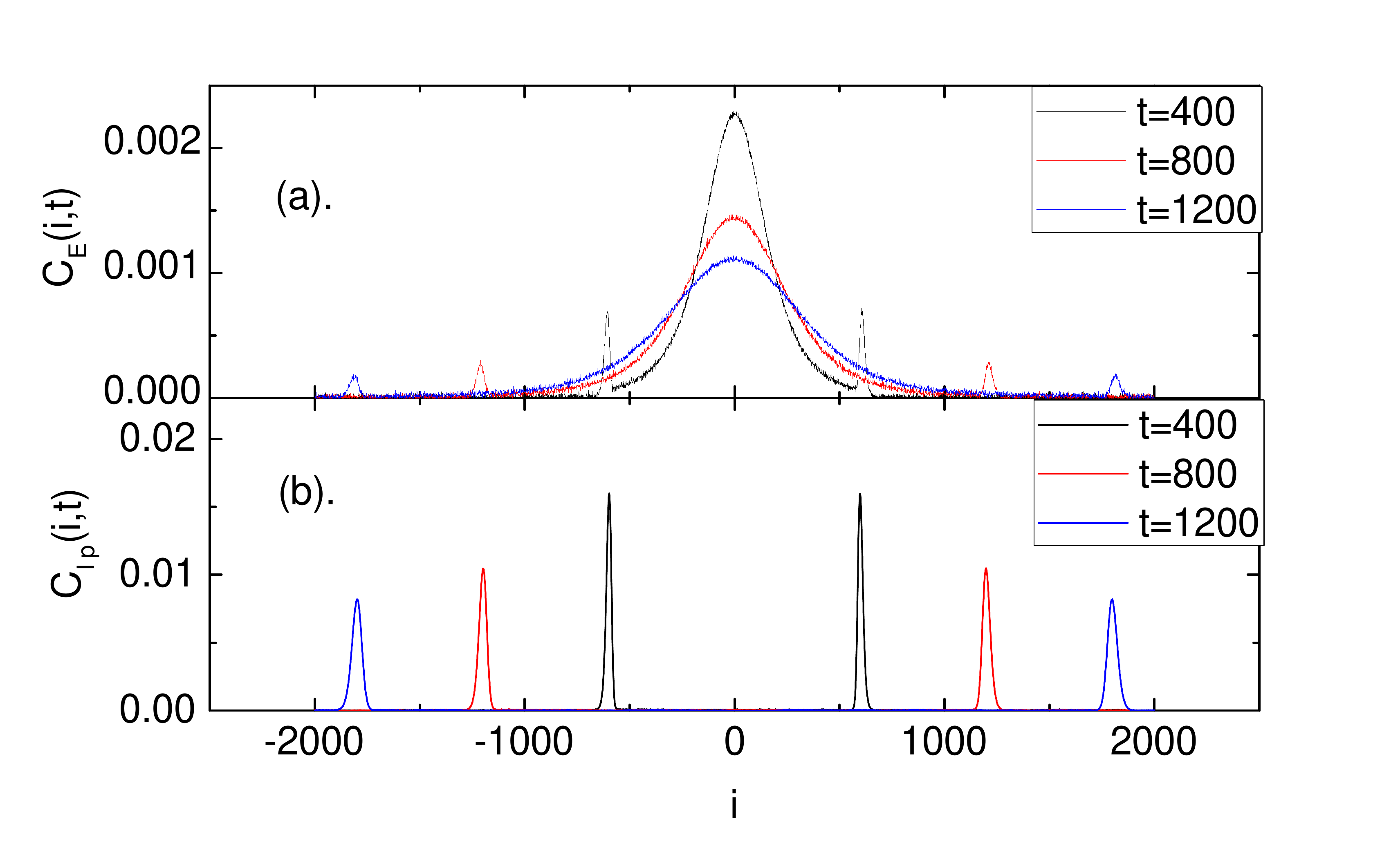}
\vspace{-0.5cm}
\caption{\label{fig:energydif-fpu}
Distribution functions $C_E(i,t)$ and $C_{I_p}(i,t)$ for energy and new conserved quantity $I_p$ which is momentum-like at three different correlation times $t=400,800$ and $1200$ for the inverse-coupling model. Lattice length is $N=4001$. The energy density $e=\langle H_i\rangle=1$ and corresponding temperature is $T=\langle p^2_i\rangle\approx 1.16$.}
\end{figure}

We then numerically study the energy diffusion behavior for the inverse-coupling model. The numerical energy diffusion method in equilibrium is proposed to calculate the spatio-temporal distribution of the energy fluctuation correlation function $C_E(i,t)$ which is defined as \cite{Zhao2006prl}:
\begin{equation}\label{eq-corr-energy}
C_E(i,t)=\frac{\langle \Delta H_i(t)\Delta H_0(0)\rangle}{\langle \Delta H_0(0)\Delta H_0(0)\rangle}+\frac{1}{N-1},
\end{equation}
where $\Delta H_i(t)=H_i(t)-\langle H_i(t)\rangle$ is the real-time energy density fluctuation at site $i$ and $\left<\cdot\right>$ means ensemble average or time average in equivalence. Here the site index $i$ is chosen from $i=-(N-1)/2$ to $(N-1)/2$ for simplicity. The extra term of constant $1/(N-1)$ is a result of using energy density instead of temperature as the input parameter in the closed system. From definition, the initial distribution is a Kronecker $\delta$ function as $C_E(i,t=0)=\delta_{i,0}$ in the thermodynamical limit $N\rightarrow\infty$. The distribution $C_E(i,t)$ describes the spatio-temporal energy spreading from the center site $i=0$ and initial correlation time $t=0$.

\begin{figure}
\includegraphics[width=1\columnwidth]{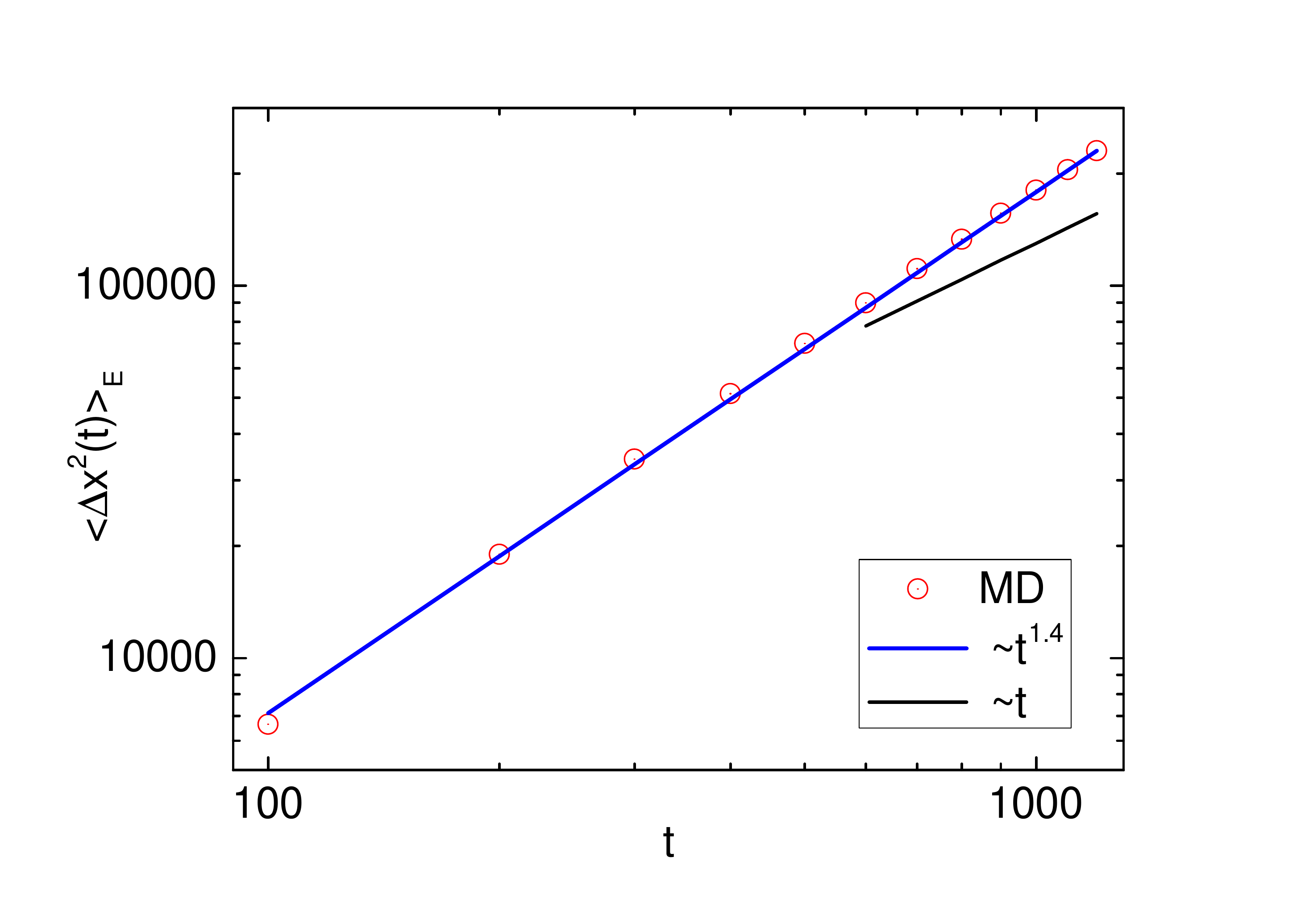}
\vspace{-0.5cm}
\caption{\label{fig:msd-fpu}
The MSD $\left<\Delta x^2(t)\right>_E$ of energy diffusion for the inverse-coupling model. The same parameters are used as in Fig. \ref{fig:energydif-fpu}. The energy diffusion is super-diffusion as $\left<\Delta x^2(t)\right>_E\propto t^{1.4}$. The curve $\sim t$ is shown for comparison.}
\end{figure}

In Fig. \ref{fig:energydif-fpu}(a), the distribution functions $C_E(i,t)$ has been plotted for an inverse-coupling model with length $N=4001$ at three different correlation times $t=400,800$ and $1200$. The energy density $e$ is set as $e=1$ which corresponds to a temperature $T=1.16$. The energy distributions $C_E(i,t)$ exhibit Levy walk distribution with two side peaks indicates anomalous diffusion, rather than normal diffusion with the Gaussian normal distribution. It is clear that these distributions are almost the same as that of FPU-$\beta$ lattice \cite{Zhao2006prl,Nianbei2010prl}. To identify the exact diffusion behavior, the MSD $\left<\Delta x^2(t)\right>_E=\sum_{i}i^2 C_E(i,t)$ has been plotted in Fig. \ref{fig:msd-fpu}. The fitted time behavior of $\left<\Delta x^2(t)\right>_E\propto t^{\beta=1.40}$ indicates that the energy diffusion in the inverse-coupling model is super-diffusion. The exponent $\beta=1.40$ is also very similar to that of FPU-$\beta$ lattice \cite{Zhao2006prl}. Although the translational symmetry and momentum conservation are broken in the inverse-coupling model, its energy diffusion does exhibit an anomalous energy super-diffusion behavior.

As we have demonstrated that for the linear inverse-coupling model, the total momentum is not conserved as $\sum_{i}dp_i/dt=-\sum_{i}(x_{i-1}+2x_i+x_{i+1})$ usually does not vanish. There is no translational symmetry as the Lagrangian of inverse-coupling model is not invariant under the transformation $x_i\rightarrow x_i + s$ with $s$ some constant. However, the Lagrangian is invariant under this transformation $h^{s}: x_i\rightarrow x_i+(-1)^{i}s$. According to Noether's theorem, one can define the following momentum-like quantity $I_p$:
\begin{equation}\label{eq-ip}
I_p=\sum_{i=1}\frac{\partial \mathcal{L}}{\partial \dot{x}_i}\frac{dh^s}{ds}=\sum_{i=1}(-1)^i p_i
\end{equation}
which is a conserved quantity. The $\mathcal{L}$ is the Lagrangian in Eq. (\ref{A4}).

\begin{figure}[t]
\includegraphics[width=1\columnwidth]{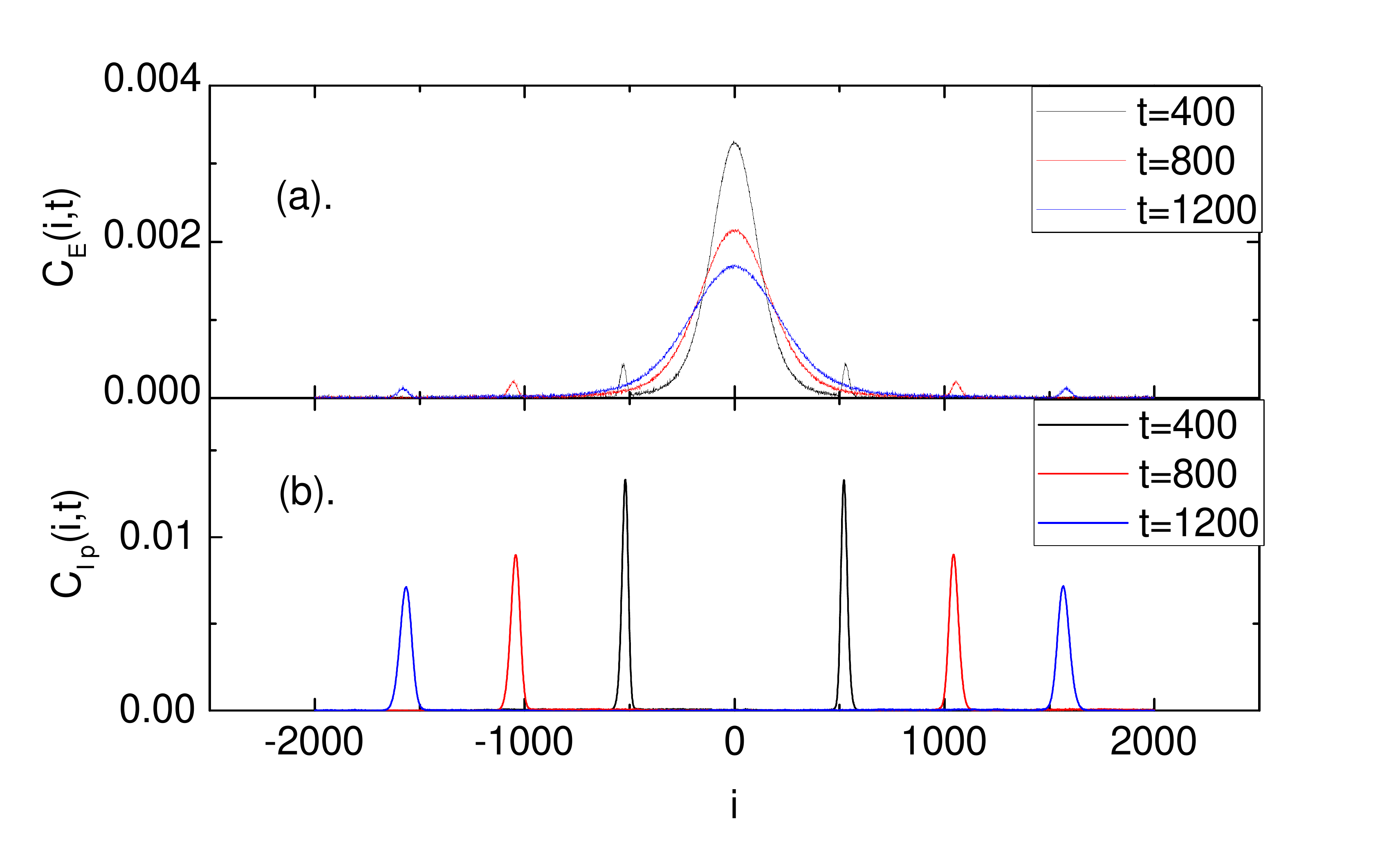}
\vspace{-0.5cm}
\caption{\label{fig:energydif-h4}
Distribution functions $C_E(i,t)$ and $C_{I_p}(i,t)$ for energy and new conserved quantity $I_p$ for high temperature limit quartic inverse-coupling model. The lattice length is $N=4001$ and the energy density is set as $e=1$ corresponding to temperature $T\approx 1.33$.}
\end{figure}

With this new momentum-like conserved quantity $I_p$, we can also calculate the distribution correlation function $C_{I_p}(i,t)=\left<(-1)^{i}p_{i}(t)p_{0}(0)\right>/T$ as we did the momentum distribution for FPU-$\beta$ lattice \cite{Zhao2006prl,Nianbei2010prl}. The spatio-temporal spreading of the $I_p$ is plotted in Fig. \ref{fig:energydif-fpu}(b) which is also the same as that for FPU-$\beta$ lattice. The new conserved quantity $I_p$ might be the reason for energy super-diffusion behavior although the zero frequency phonon mode at $k=\pi$ is not the long-wave length limit phonon with $k=0$.

To eliminate the temperature influence for the energy diffusion behavior, we also study the high-temperature limit inverse-coupling model with pure quartic interaction term in the Hamiltonian:
\begin{equation}
H=\sum_{i}\left[\frac{p_i^2}{2}+\frac{1}{4}(x_{i}+x_{i-1})^4\right],
\end{equation}

According to the renormalization phonon theory \cite{Alabiso1995jsp,Alabiso2001jpa,Lepri1998pre,Nianbei2006epl,Dahai2008pre,Nianbei2012aa,Liu2015pre}, this quartic inverse-coupling model has also renormalized phonon dispersion relation as $\omega^R_k=2\sqrt{\alpha}\cos{(k/2)}$ where the derived coefficient $\alpha$ is the same as that for FPU-$\beta$ lattice. We have numerically verified that the renormalized phonon frequency lies between the predictions of lower and upper limits represented by $\alpha_L$ and $\alpha_U$. For the quartic inverse-coupling model, the calculated $\omega^R_k$ is close to the prediction of upper limit (not shown here).

\begin{figure}[t]
\includegraphics[width=1\columnwidth]{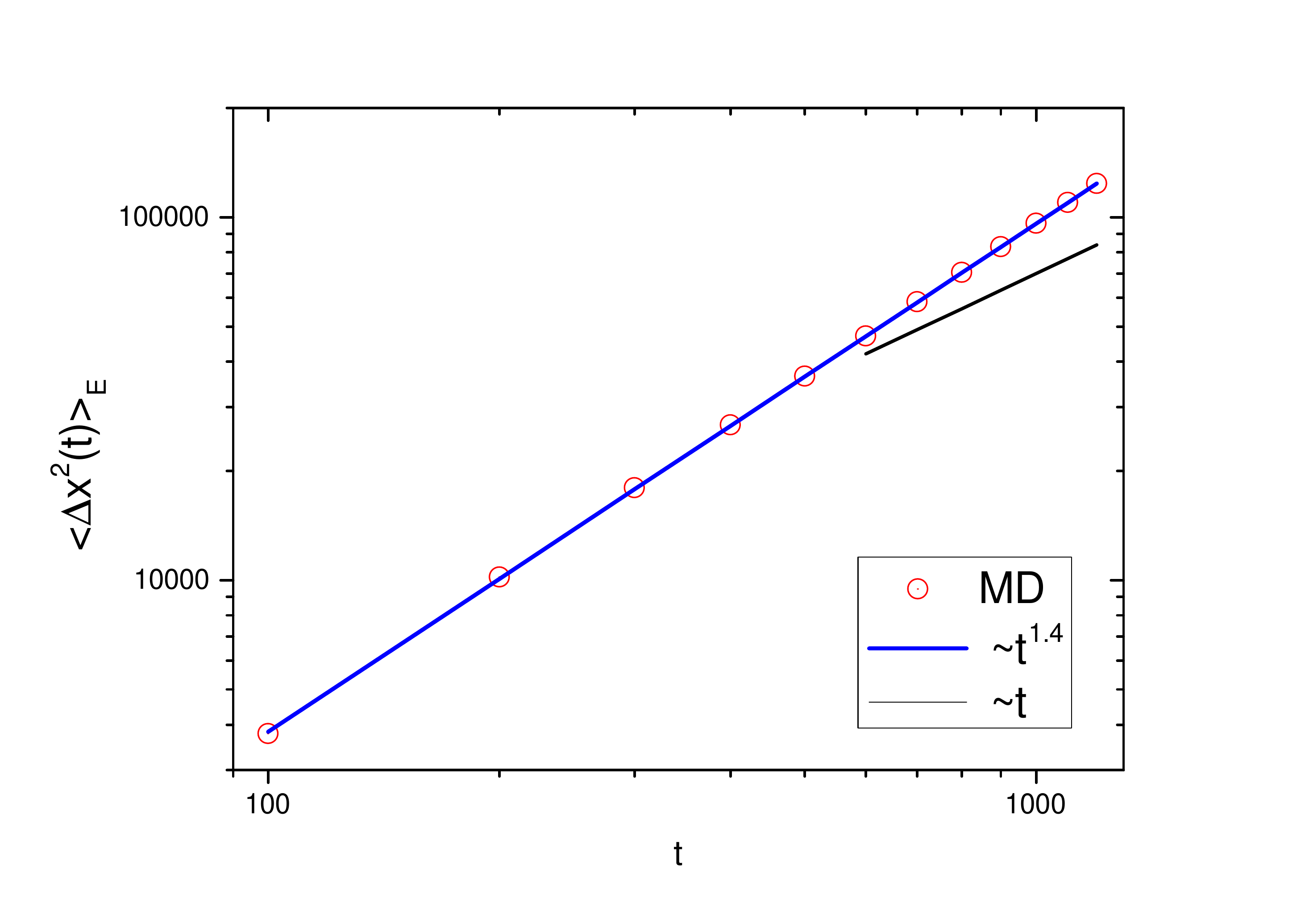}
\vspace{-0.5cm}
\caption{\label{fig:msd-h4}
The MSD $\left<\Delta x^2(t)\right>_E$ of energy diffusion for the quartic inverse-coupling model. The same parameters are used as in Fig. \ref{fig:energydif-h4}. The energy diffusion is still super-diffusion as $\left<\Delta x^2(t)\right>_E\propto t^{1.4}$. The curve $\sim t$ is shown for comparison.}
\end{figure}

In Fig. \ref{fig:energydif-h4}, both the distributions for energy of $C_E(i,t)$ and momentum-like quantity $I_p$ of $C_{I_p}(i,t)$ at three different correlation times $t=400,800$ and $1200$ are plotted for the quartic inverse-coupling model. The size is $N=4001$ and the energy density $e=1$ with corresponding temperature $T=1.33$. The quartic inverse-coupling actually denotes the high temperature or strong nonlinearity limit of the inverse-coupling model. It can be seen that the energy distribution $C_E(i,t)$ still shows a Levy walk distribution which is a signature of energy super-diffusion. The $I_p$ distribution $C_{I_p}(i,t)$ also has two ballistic wave fronts just as the momentum distribution of FPU-$\beta$ lattice. The MSD $\left<\Delta x^2(t)\right>_E$ for energy fluctuation is plotted in Fig. \ref{fig:msd-h4}, and a super-diffusion with $\left<\Delta x^2(t)\right>_E\propto t^{\beta=1.40}$ is obtained.

In conclusion, we have proposed a 1D inverse-coupling model without translational symmetry. The total momentum is not conserved any more while the zero frequency phonon mode is maintained as there is no on-site potential. Our numerical results show that this momentum non-conserving inverse-coupling model exhibits energy super-diffusion behavior corresponding to anomalous heat conduction. Therefore the momentum non-conservation is not the sufficient condition for normal energy diffusion or heat conduction in 1D non-integrable lattices.
However, our proposed model indeed has zero frequency phonon mode as a result of lacking on-site potential. This leaves the claim that on-site potential is a sufficient condition for normal heat conduction in 1D nonlinear lattices still valid.

This work is supported by NSFC with grant No. 11775158, No. 11775159, the Science and Technology Commission of Shanghai Municipality with grant No. 17ZR1432600, No. 18ZR1442800, No. 18JC1410900, the Opening Project of Shanghai Key Laboratory of Special Artificial Microstructure Materials and Technology, and the Scientific Research Funds of Huaqiao University.

\appendix
\section{The Hamiltonian of inverse coupling model}

As shown in Fig.\ref{fig:pic1}, supposing $\theta_i$ is always small and taking the limit $r/L\rightarrow0$, the governing equation of $x_i$ is:
\begin{equation}
\begin{split}
\ddot{x}_i+\gamma(2x_i-r\theta_{i-1}+r\theta_i)\\+\beta(x_i-r\theta_{i-1})^3+\beta(x_i+r\theta_i)^3=0.
\end{split}
\label{A1}
\end{equation}

Supposing that the rotational inertia of pole is small and its kinetic energy is ignorable compared to its potential, we have:
\begin{equation}
\begin{split}
\gamma(x_i-r\theta_{i-1})+\beta(x_i-r\theta_{i-1})^3\\-\gamma(x_{i-1}+r\theta_{i-1})-\beta(x_{i-1}+r\theta_{i-1})^3=0,
\end{split}
\label{A2}
\end{equation}
which yields $2r\theta_{i-1}=x_i-x_{i-1}$.
Analogously, $2r\theta_{i}=x_{i+1}-x_{i}$. Substituting the expression of $\theta_{i-1}$ and $\theta_{i}$ into Eq.(\ref{A1}), we get
\begin{equation}
\begin{split}
\ddot{x}_i+\frac{\gamma}{2}(2x_i+r\theta_{i-1}+r\theta_i)\\+\frac{\beta}{8}(x_i+r\theta_{i-1})^3+\beta(x_i+r\theta_i)^3=0.
\end{split}
\label{A3}
\end{equation}

Eq.(\ref{A3}) is equivalent to the Lagrange Equation of a system with Lagrangian:
\begin{equation}
\mathcal{L}=\sum_i\left(\frac{\dot{x}_i^2}{2}-\frac{\gamma'}{2}(x_i+x_{i-1})^2-\frac{\beta'}{4}(x_i+x_{i-1})^4\right),
\label{A4}
\end{equation}
where $\gamma'=\gamma/2$ and $\beta'=\beta/8$. For simplicity, we set $\gamma'=\beta'=1$.

Namely, the system is reduced into one with only N degrees of freedom since the Lagrangian is independent of $\theta_i$ . And the Hamiltonian in Eq.(\ref{ic}) is obtained just by taking a Legendre transform for the Lagrangian in Eq.(\ref{A4}).

\bibliographystyle{apsrev4-1}

\end{document}